# A Fully Digital Relaxation-Aware Analog Programming Technique for HfOx RRAM Arrays

Hamidreza Erfanijazi, Luis A. Camuñas-Mesa, Elisa Vianello, Teresa Serrano-Gotarredona *Senior Member, IEEE*, and Bernabé Linares-Barranco, *Fellow, IEEE*

***Abstract*— For neuromorphic engineering to emulate the human brain, improving memory density with low power consumption is an indispensable but challenging goal. In this regard, emerging RRAMs have attracted considerable interest for their unique qualities like low power consumption, high integration potential, durability, and CMOS compatibility. Using RRAMs to imitate the more analog storage behavior of brain synapses is also a promising strategy for further improving memory density and power efficiency. However, RRAM devices display strong stochastic behavior, together with relaxation effects, making it more challenging to precisely control their multi-level storage capability. To address this, researchers have reported different multi-level programming strategies, mostly involving the precise control of analog parameters like compliance current during write operations and/or programming voltage amplitudes. Here, we present a new fully digital relaxation-aware method for tuning the conductance of analog RRAMs. The method is based on modulating digital pulse widths during erase operations while keeping other parameters fixed, and therefore requires no precise alterations to analog parameters like compliance currents or programming voltage amplitudes. Experimental results, with and without relaxation effect awareness, on a 64 RRAM 1T1R HfOx memory array of cells, fabricated in 130nm CMOS technology, indicate that it is possible to obtain 2-bit memory per cell multi-value storage at the array level, verified 1000 seconds after programming.**

***Index Terms*— Analog RRAM, Multi-Level RRAM, Pulse width modulation, Conductive states, Memristive crossbar, Relaxation and retention effects, Neuromorphic applications.**

## I. INTRODUCTION

HfOx-based Resistive Random-Access Memory (RRAM) technology is promising for implementing low-power non-volatile, compact but computationally-efficient CMOS-compatible neuromorphic edge devices. RRAM devices can be used to emulate the synaptic knowledge-storage devices in the brain which, being connected between neurons, also directly participate in computing processes, overcoming the memory-wall power problem of Von-Neumann-based computing systems [1]. This efficiency can be further boosted if RRAM devices can store multi-level values. For example, brain synapses have been shown to be capable of storing 26 different weights each (4.7 bits per synapse) [2]. In practice, however, RRAM devices suffer from considerable device-to-device and cycle-to-cycle variability [3]-[4], short-term and long-term stochastic conductance drift effects, and aging [3]-[4], and are therefore mainly used reliably as binary (2-state) memory devices (1 bit per synapse) [8]. Nonetheless, some techniques have been reported recently which are capable of successfully achieving multi-level storage per individual RRAM device in monolithically fabricated CMOS+RRAM crossbar 1T1R arrays (see Fig. 1). The reported techniques typically require iterative write/erase steps per device by gradually precision tuning a given analog parameter per each 1T1R device. Some researchers, for example, report having achieved successful 4-level post-relaxation storage by careful iterative tuning of the analog compliance current at each write operation [9]-[14]. Another technique consists of progressively changing analog write and/or erase voltage pulse amplitudes [15]-[21]. The added problem of stochastic degradation of the analog programmed value seems to comprise two components: (a) a fast sub-5-second and stronger short-term relaxation, and (b) a slower minute-range but weaker long-term retention loss [17]-[18]. To mitigate these effects, Esmanhotto et al. [10] proposed introducing a 5-second waiting step at device level before every read verification after each programming attempt, achieving 4-level programming after long-term degradation stabilization. Alternatively, at system level, short- and long-term degradation effects could be considered and modeled during network training, resulting in an overall system tolerant to such stochastic drifts [17]-[18]. Chen et al. [19] proposed a new device based on HfOx atomic layer deposition with post-metal annealing. By applying identical pulses with reset-verification (IPRV) they demonstrate degradation-free conductance states.

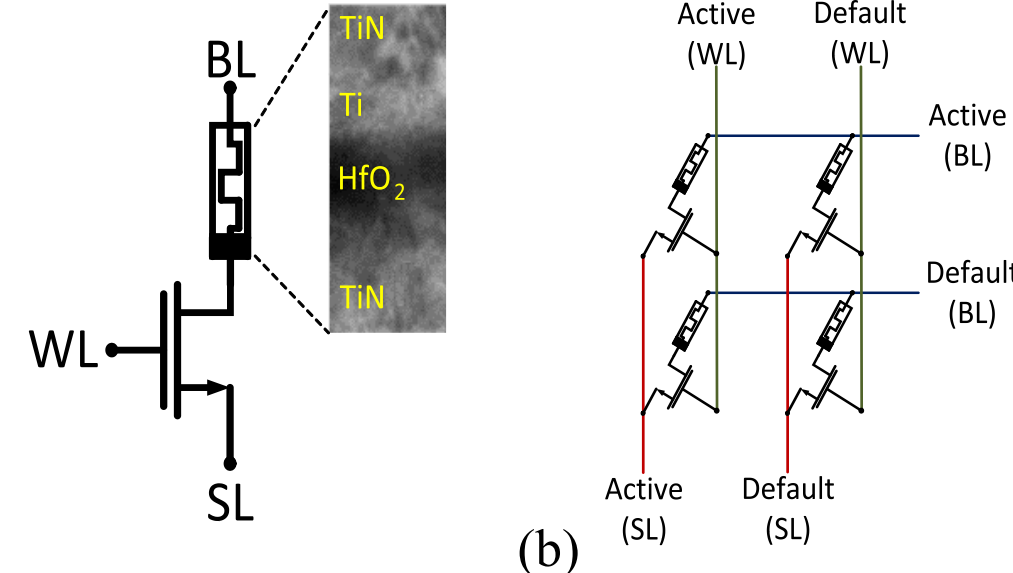

Fig. 1.(a) Schematic diagram of a 1T1R memory (synaptic) element. (b) Representation of a 1T1R crossbar: "Default" column and row lines represent multiples sharing same WL, BL,SL voltages, while there are only one "Active" (WL, SL) column and line (BL) selecting one single 1T1R memory device.

In this brief we present an experimentally-verified fully-digital device-level, short-term relaxation-aware, iterative, multi-level programming technique which does not require the careful tuning of analog parameters during each device write and/or erase iteration, for the standard HfOx devices in Fig. 1(a). Fig. 2 illustrates typical hysteresis behavior during write/erase sweeps in such HfOx 1T1R devices [22]-[23]. At increasing positive voltages ($V_{BL} - V_{SL} > 0$), devices tend to transition abruptly from a low conductance state to a compliance-current-controlled higher conductance state (*Write* operation). On the other hand, with negative voltages ($V_{BL} - V_{SL} < 0$) they tend to transition smoothly from higher to lower continuous conductance levels (*Erase* operation). This observation inspired

This work was partly funded by EU grants 871501, 824164, 899559, 101070908, and National grant PID2019-105556GB-C31.

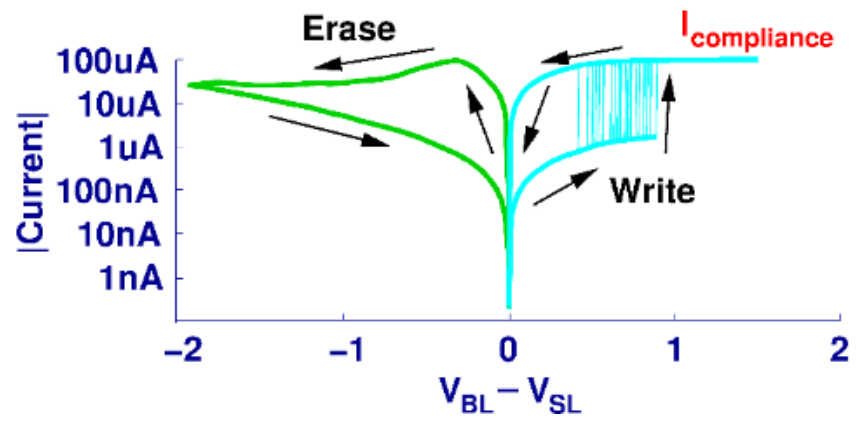

Fig. 2. Typical I/V write/erase hysteresis behavior

us to use only erase operations to seek the target analog programming level. Also, since we did not want to change the analog amplitude of the erasing pulses, we instead tried progressively tuning the erasing pulse width digitally. In the technique proposed here, we therefore performed *Write* operations with fixed compliance current and fixed voltage pulse width and amplitude, combined with *Erase* operations with fixed amplitude voltage pulses but variable, digitally-controlled widths. The experimental results were obtained on a 64 1T1R crossbar array fabricated in 130nm CMOS with BEOL HfOx RRAM devices. Programming without relaxation effect awareness, we obtained verified 3-level conductance 1000s after the array programming was complete, whereas programming with short-term relaxation effect awareness we obtained 4-level conductance.

Table I. *Active* and *Default* mode voltages for the crossbar lines in each operation, and pulse widths for the corresponding *Active* modes

| Operation | Mode | WL | SL | BL | Pulse Width |
|---|---|---|---|---|---|
| Form | Active | 1.55V | 0 V | 4.8 V | 40µs |
| | Default | 0 V | 4.8 V | 2.4 V | Fixed |
| Write | Active | 1.24 V | 0 V | 2.4 V | 100ns |
| | Default | 0 V | 2.4 V | 2.4 V | Fixed |
| Erase | Active | 4.05 V | 1.07 V | 0 V | CP x 10ns |
| | Default | 0 V | 0 V | 2.4 V | Fixed |
| Read | Active | 3.38 V | 2.1 V | 2.4 V | 200µs |
| | Default | 0 V | 2.4 V | 2.4 V | Fixed |
| Stand-by | - | 0 V | 2.4 V | 2.4 V | Fixed |

## II. PROPOSED PROGRAMMING TECHNIQUE

Fig. 1(b) schematically represents an arbitrary-size crossbar of 1T1R devices. One column and one row of 1T1R devices is set to the *Active* mode. The other columns and rows are set to *Default* mode. Table I shows the configurations of the 1T1R devices in the crossbar for the different operations and modes, indicating the voltage values at the word lines (WL), bit lines (BL) and source lines (SL). In the *Default* mode, voltages are set to prevent any changes in the memristor states. In the *Active* modes, the corresponding lines (WL, BL, SL) are set to activate only one 1T1R target device: the one at the intersection of the horizontal and vertical *Active* lines. To access this target device, its SL and BL lines are first changed to the corresponding *Active* values, and then, during the time shown in the *Pulse Width* column in Table I, its gate line WL is set to the *Active* voltage value. For the NMOS transistor selector in Fig. 1 and for the peripheral driving circuitry, thick oxide transistors were used that allowed for a power supply voltage of up to $V_{DD}$ = 4.8V. Table I also shows a *Stand-by* state in which the SL and BL lines of all columns and rows are set to $V_{DD}/2$ while all gate lines are set to 0V. In this *Stand-by* state, there is no active line or column and no 1T1R device is targeted. The *Stand-by* state is used to transition safely between operations, avoiding possible glitches at the different lines that could unintentionally change the memory state of the memristors in the crossbar. Initially, all crossbar devices were formed with 40µs full $V_{DD}$ pulses and a compliance current of about 600µA [8]. The write operation is always a full write, for which a write pulse of $V_{BL} - V_{SL} = 2.4V$ is applied to the target 1T1R device for 100ns while its compliance current is set to about 300µA by setting the gate-to-source voltage of the NMOS selector to 1.24V. For erase operations, erase pulse widths are iteratively and digitally adjusted to a multiple of 10ns (CP x 10ns, where CP is an integer). For the erase pulse amplitude $V_{SL} - V_{BL}$ and gate voltage $V_{WL}$ a judicious compromise was reached after many trials to optimize the post-relaxation multi-level storage, resulting in the chosen values of 1.07V and 4.05V, respectively, as shown in Table I. All these voltage values were kept fixed for all devices and all programming iterations. The sequence of operations and variable-width programming iterations are controlled by an external FPGA with a 100MHz clock frequency. In future versions, we intend to use an on-chip programmable-width pulse generator with sub-ns width control [24]. Fig. 3 shows a simplified diagram of the crossbar and Active/Default selector circuits in the fabricated chip. For each Active/Default selector block, the external FPGA selects the *Active* line through the digital control word $SEL_{XX}$ (XX = WL, SL, or BL), and connects it to an external *Active* voltage bias line. The other lines are connected to an external *Default* voltage bias line. Lines $Enable_{SL}$ and $Enable_{BL}$ are normally kept ON, while line $Enable_{WL}$ is set to ON only during pulse duration. The active WL and SL lines are directly connected to external lines $V_{WLactive}$ and $V_{SLactive}$, which are in turn connected by the FPGA to the pre-set bias voltages shown in Table I, depending on the operation. For a *Read* operation, the selected *Active* BL line is connected to the external opamp-based current sensing circuit shown on the right of Fig. 3. For other operations (not shown in Fig. 3), it is connected directly to bias voltage $V_{BLactive}$. During a *Read*, the selected *Active* 1T1R device current $I_R$ is detected by sensing resistor $R_{sense}$. This resistor's voltage $V_R = R_{sense} \times I_R$ is measured by a DAC run by the FPGA. This way, the target 1T1R device conductance is obtained from the following equation

$$G_{1T1R} = \frac{V_R / R_{sense}}{V_{BL} - V_{SL}} \quad (1)$$

For multi-level conductance programming, the conductance range is divided into a number of intervals, each with a low value $G_L$ and a high value $G_H$, inside which the conductance will be programmed. Consecutive intervals are assigned with a conductance gap in between, to safeguard against post-programming drifts, regardless of whether a relaxation-aware technique is being used. Different ways of assigning conductance intervals were proposed in previous works. Here we use an empirical mixture of both linear and sigma-based techniques [9], as shown later in Section III.

We compared two programming techniques: one that ignores relaxation effects and another that considers short-term relaxation of the programmed weights:

### *A. Programming without Relaxation Awareness:*

First, we programmed our crossbar array without considering relaxation effects. The algorithm used, implemented on an auxiliary FPGA, is shown in Fig. 4. Low $G_L$ and high $G_H$ values were set for each programming interval $n$, the erase pulse width coefficient *CP* was initialized to zero, and the target 1T1R conductance $G$ was read. After each *Read*, if $G$ lay inside the target programming interval, programming was finished. If it was above ($G > G_H$) we tried to lower $G$ by strengthening the next *Erase* operation and increasing coefficient *CP* by one unit. We then performed the *Erase* operation and again read the new $G$. If, after a *Read*, the conductance $G$ was below the target interval ($G < G_L$), we performed a full write with the pulse width fixed at 100ns, at the same time reducing coefficient *CP* by 1 in order not to strengthen the next *Erase* operation. Fig. 5 illustrates this procedure for one particular device programming sequence. The example target interval, shaded in gray, is given by $G_L = 33.2\mu S$ and $G_H = 38.08\mu S$. The horizontal axis represents the programming iteration number. The left vertical axis represents the read conductance $G$, and the right vertical axis represents the erase pulse width (*CP* x 10ns, in red circles) or write pulse width (100ns, in blue squares) used immediately before the corresponding read value $G$. In this example, there are a total of 16 write/erase iterations until a $G$ value inside the target interval is read. Since this conductance interval is in the lower range, most of the read conductance values are above it, resulting in the execution of *Erase* operations with increasing pulse widths (except in iteration 5, where a *Write* operation was performed). In iteration 4, an *Erase* operation was performed after which the read $G$ was below the target interval. *CP* was therefore decreased, followed by a *Write* in iteration 5. In this iteration the newly read $G$ value was above the interval, thus increasing CP again and providing an *Erase* pulse for iteration 6 of the same width as the latest *Erase* pulse (iteration 4). It is interesting to observe the stochastic nature of the different *Erase* pulses, as they produced both increments and decrements in conductance. Although there did not appear to be any clear correlation between pulse width and conductance, such correlation did in fact exist, as described in Section III.

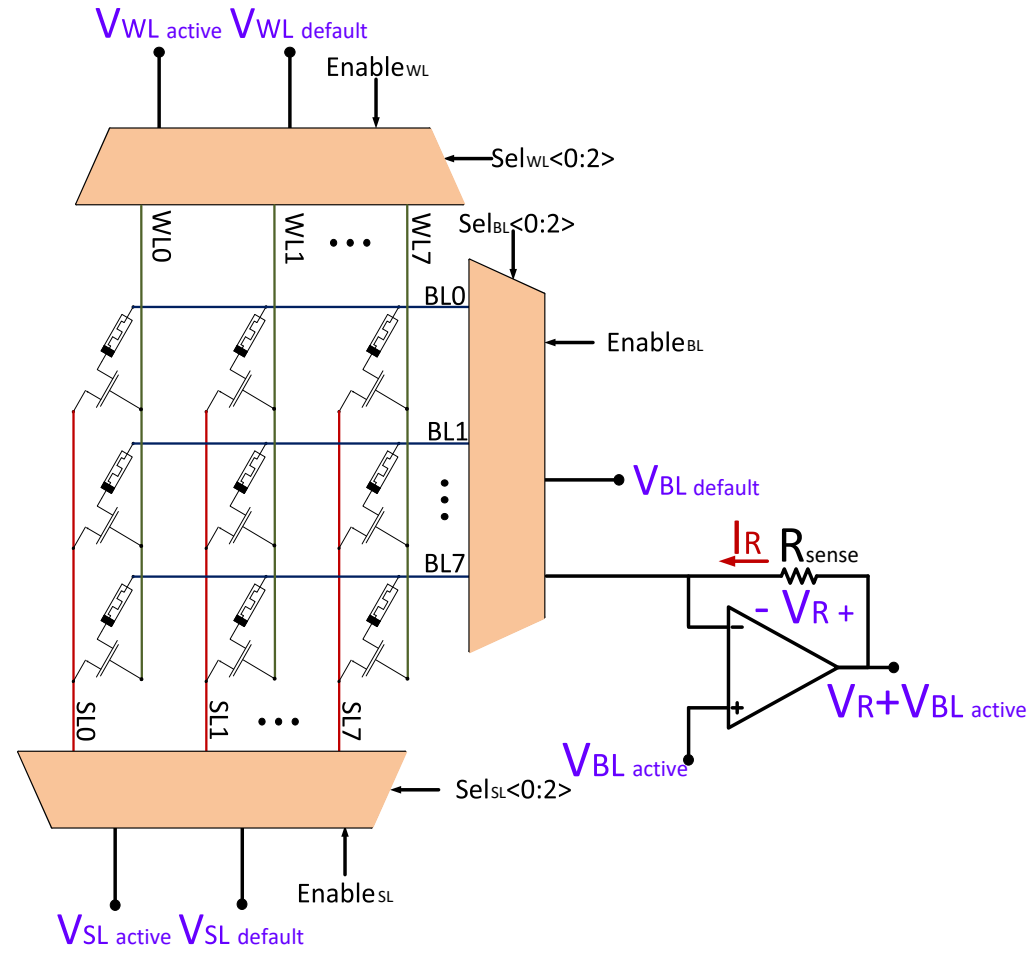

Fig. 3. Schematic architecture of the 8x8 1T1R crossbar.

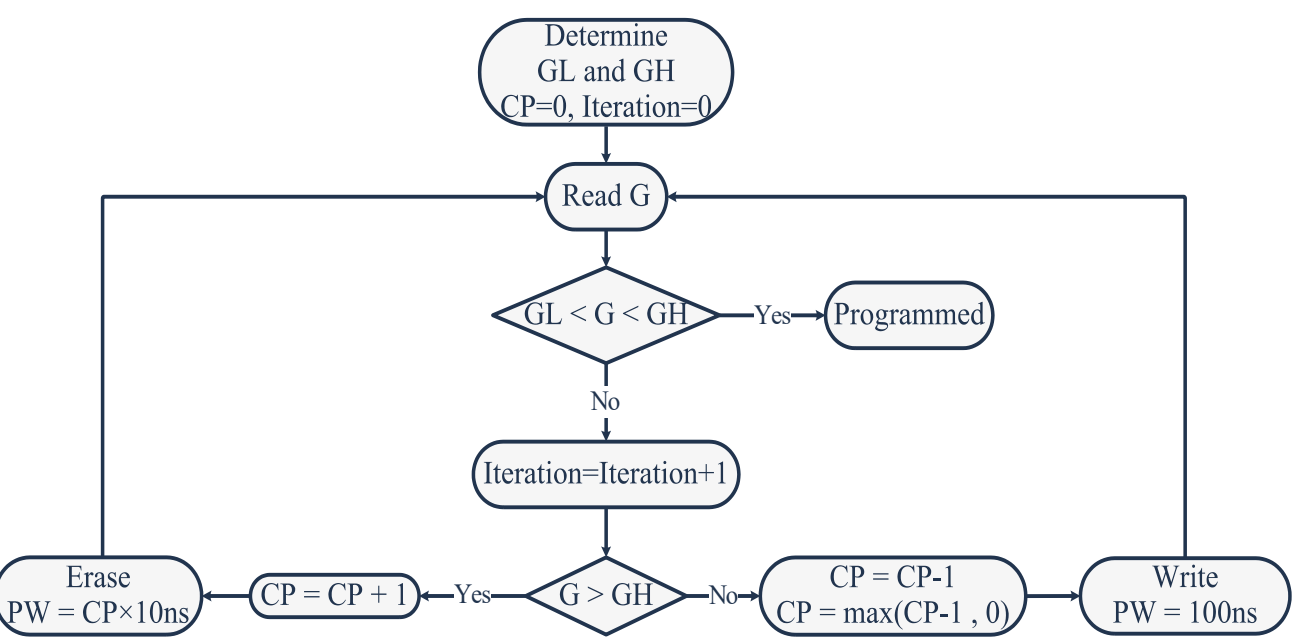

Fig. 4. Multi-level programming algorithm without relaxation awareness

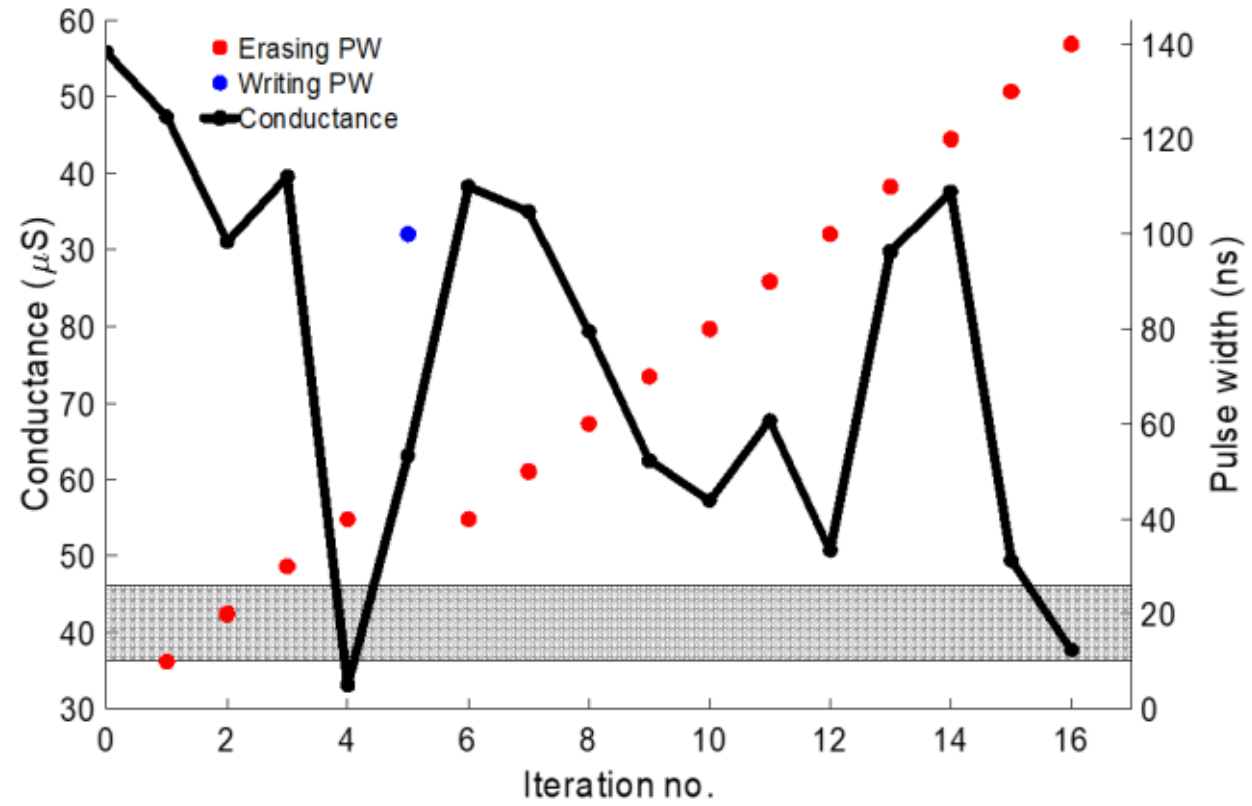

Fig. 5. Experimental illustration of a programming sequence for the algorithm in Fig. 4.

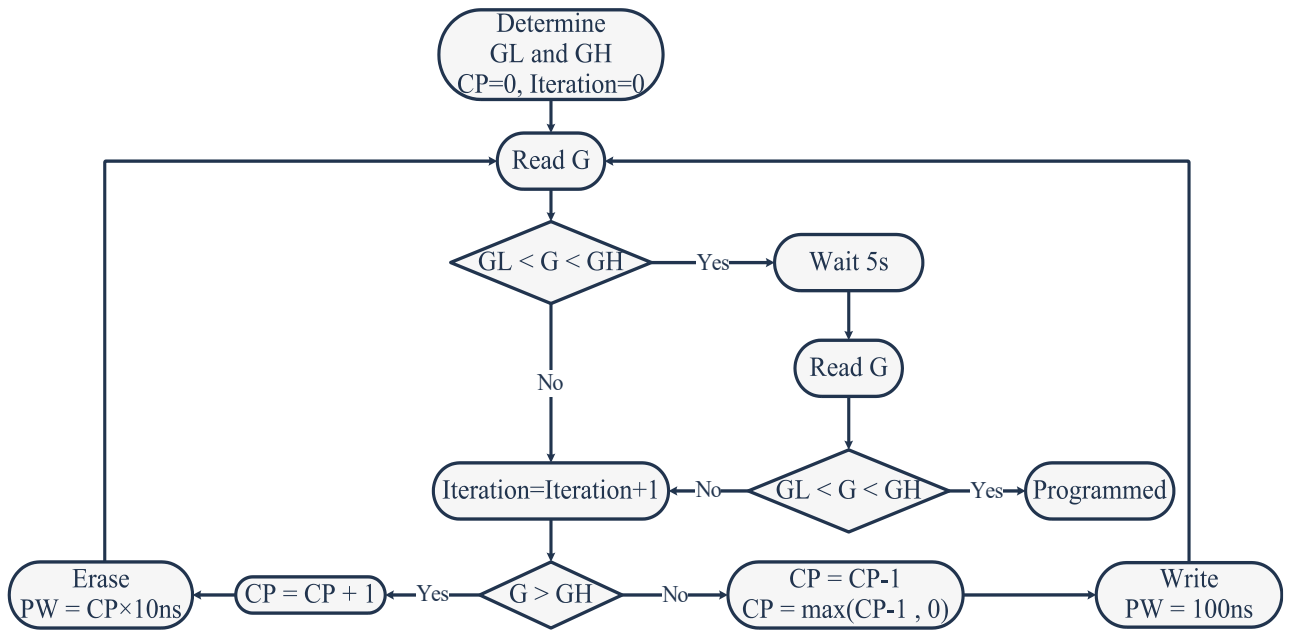

Fig. 6. Multi-Level programming with short-term relaxation awareness

*B. Programming with Short-Term Relaxation Awareness:*

It is well known that sub-micron-size HfOx 1T1R devices suffer from short- and long-term stochastic degradation effects, which are more pronounced a) when trying to program multi-level conductance levels, and b) during the first few seconds after programming [9],[17]. To compensate for this, it is proposed [10] waiting for a time $\Delta t$ after each programming before doing any *Read* operations, concluding that with $\Delta t$ = 5s four distinct conductance levels could safely be programmed in a crossbar array. Inspired by this mitigation technique, which we verified experimentally, here we propose a slightly modified alternative. After each programming step, we perform a *Read* operation without waiting. If the conductance value read lies outside the target interval, we proceed as in Fig. 4; but if the conductance value read lies inside the target interval, we wait for time $\Delta t$ and then perform a second *Read*. If, after this second *Read with waiting*, the conductance still lies inside the target interval, we consider the device as programmed, otherwise the programming procedure continues. This is illustrated in Fig. 6.

## III. Experimental Results

A monolithic CMOS+RRAM test chip was fabricated in CEA-LETI 130nm technology, including an 8x8 1T1R HfOx-based crossbar array, as illustrated in Fig. 3. Fig. 7 shows the experimental test setup including the CMOS-RRAM chip PCB and the FPGA control PCB that implements the algorithms in Figs. 4 or 6. The target conductance range for programming multi-level values was selected as about [0, 100μS], as a good compromise between number of programming levels while avoiding excessive power dissipation. For this range, the value used for the sense resistor in Fig. 3 was $R_{sense} = 20K\Omega$. After exhaustive experimentation and trials, we opted for an initial allocation of conductance intervals which were like a mixture of the linear- and sigma-based conductance allocation methods [9] and are as shown in Table II. The optimum number of initial programming intervals was found to be 8, denominated $n = i$, with $i = 0, \ldots 7$. Increasing this number resulted in failure of programming convergence for many memristors.

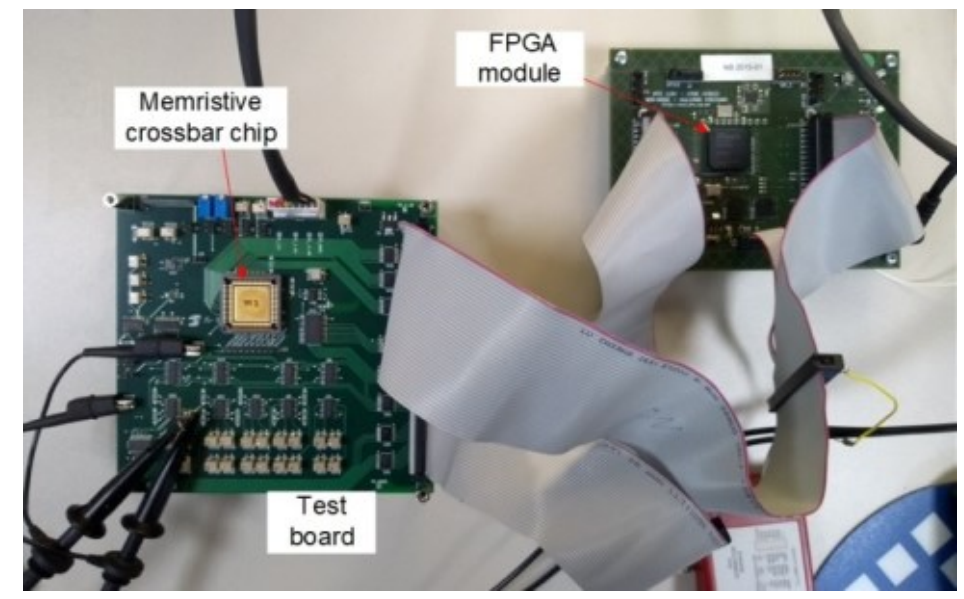

Fig. 7. Test measurement setup including a PCB hosting the 1T1R crossbar chip and an external FPGA control PCB

Table II . Initial eight-level conductance interval allocation

| State# | Conductance interval (μs) | $G_L$(μs) | $G_H$(μs) |
|---|---|---|---|
| n=0 | - | 0 | 30 |
| n=1 | 4.88 | 33.2 | 38.08 |
| n=2 | 3.3 | 41.3 | 44.6 |
| n=3 | 3.3 | 47.8 | 51.1 |
| n=4 | 3.3 | 52.7 | 56 |
| n=5 | 3.2 | 57.6 | 60.8 |
| n=6 | 1.6 | 64.1 | 65.7 |
| n=7 | - | 71.2 | 100 |

Fig. 8 shows the experimentally obtained results when using the faster programming algorithm in Fig. 4, which always performed a *Read* operation immediately after a *Write* or *Erase* operation, without any waiting interval. The horizontal axis in Fig. 8 represents the initially programmed conductance values $G_{init}$ for each of the 64 1T1R devices and for each of the initial target states, $n = 0, \ldots 7$. The vertical axis represents, for the same devices, the $G_{1ks}$ values measured 1000 seconds after full array programming, which we observed is a conservative waiting interval after which negligible degradation was detected. The black solid-line square boxes indicate the initial target intervals in Table II for both axes. It can be seen that the lowest ($n = 0$) and highest ($n = 7$) conductance states underwent little or negligible drift after 1ks. The intermediate conductance states, however, underwent considerable drift, which seemed to be more pronounced and upward for the lower states ($n = 1, 2, 3$ and 4), while the higher states ($n = 5, 6$) seemed to undergo a more symmetric upward and downward drift. Looking at the final $G_{1ks}$ values in Fig. 8, it can be concluded that there are only

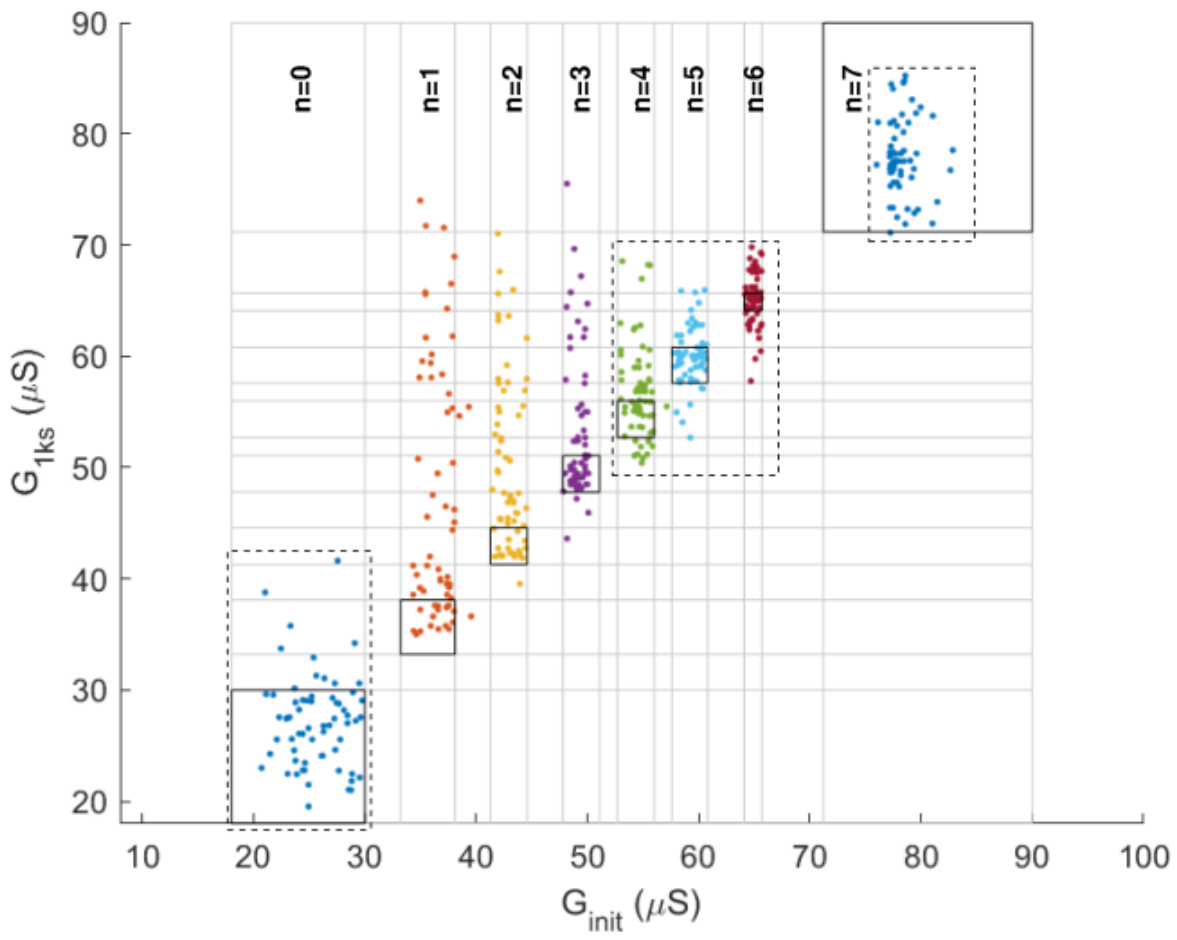

Fig. 8. Final conductance values measured for the 64 devices 1ks after programming when using the faster programming technique in Fig. 4, which neglected relaxation effects.

3 states that can be distinguished without overlap: $n = 0$, $n = 7$, and one of either $n = 4$, 5, or 6. This suggests that the same 3-level result can be obtained by removing states $n = 1$, 2, and 3 in Table II, and collapsing states $n = 4$, 5, and 6 into one single state with $G_L = 52.7$μS and $G_H = 65.7$μS. The three distinguishable states are indicated by the segmented-line boxes.

Fig. 9 shows the experimentally measured results when using the short-term relaxation aware technique described in Fig. 6 with a waiting time of $\Delta t = 5$s to ensure the conductance remains within the target interval. For longer waiting times, the result was similar, and for shorter ones slight degradations were observed. 5s was therefore a good compromise. It can be seen there are now 4 non-overlapping states after 1000 seconds, $n = 0$, 3, 6, 7. In Fig. 9 the short-term relaxation effects (at least those slower than 5s) are removed. Some residual, slower and more symmetric relaxation effects still remain, however, especially for states $n = 1$ to 4. Note that now, it is not necessary to collapse states but simply to limit the initial target states to $n = 0, 3, 6, 7$ in Table II in order to obtain distinguishable 4-level conductance intervals. In Fig. 5 we saw that the programing process had a high degree of stochasticity. To find out whether any correlation exists between the target states and the *Final Erase Pulse Width* (FEPW), Fig. 10 shows, for each 1T1R device and each initial target state defined in Table II ($n = 0$ to 7), the FEPW used for all programming iterations and for both programming algorithms. A strong correlation can be seen: the average FEPW increases as target conductance values decrease. However, there is also a strong stochasticity, as revealed by the high values of the standard deviations for the FEPWs. Table III shows, for each target state, the average number of programming iterations used ($\overline{n_{iter}}$), the average, and the standard deviation of FEPW for each programming cycle and for both algorithms. For $\overline{n_{iter}}$ there is no clear correlation with the target state. The $\Delta\overline{n_{iter}}$ column represents the average number of iterations the algorithm in Fig. 6 went through the $\Delta t$ branch. Here too, there is a clear correlation: as the conductance increases, the algorithm requires more times to go through the waiting branch. An iteration without waiting takes about 120ms, mainly due to settling delays when changing operation modes. Also note in Table III that the slow method requires slightly more iterations, mainly because short-term drift between iterations makes convergence more difficult.

Table III . Average iterations, average FEPW, and σ(FEPW)

| | No waiting (Fig. 4, Fig.10a) | | | $\Delta t = 5$s (Fig. 6, Fig. 10b) | | | |
|---|---|---|---|---|---|---|---|
| State# | $\overline{n_{iter}}$ | FEPW (ns) | σ(FEPW) (ns) | $\overline{n_{iter}}$ | FEPW (ns) | σ(FEPW) (ns) | $\Delta\overline{n_{iter}}$ |
| n=0 | 19 | 184 | 85.1 | 21 | 193 | 153.3 | 2 |
| n=1 | 10 | 150 | 107.6 | 13 | 161 | 164.3 | 3 |
| n=2 | 9 | 105 | 93.8 | 12 | 86.6 | 122.4 | 3 |
| n=3 | 15 | 78.0 | 49.1 | 21 | 78.8 | 56.3 | 6 |
| n=4 | 14 | 63.3 | 39.3 | 20 | 61.6 | 36.9 | 6 |
| n=5 | 13 | 42.7 | 16.6 | 24 | 39.1 | 14.11 | 11 |
| n=6 | 11 | 26.1 | 9.3 | 16 | 30.8 | 2.7 | 5 |
| n=7 | 3 | 27.8 | 19.4 | 26 | 22.2 | 16 | 23 |

Table IV Comparison with other relaxation-aware methods

| Reference | Device | Method | Intervals (μS) | States | Non-overlapping states |
|---|---|---|---|---|---|
| This work | $HfO_2$/Ti | PWM | 0-100 | 4 | Yes |
| [10] | $HfO_2$ | CC | 25-130 | 4 | Yes |
| [16] | $TiO_2$ | PAM | 232-500 | 4 | Yes |
| [22] | $HfO_2$/Ti | PAM | 0-500 | 15 | NO |

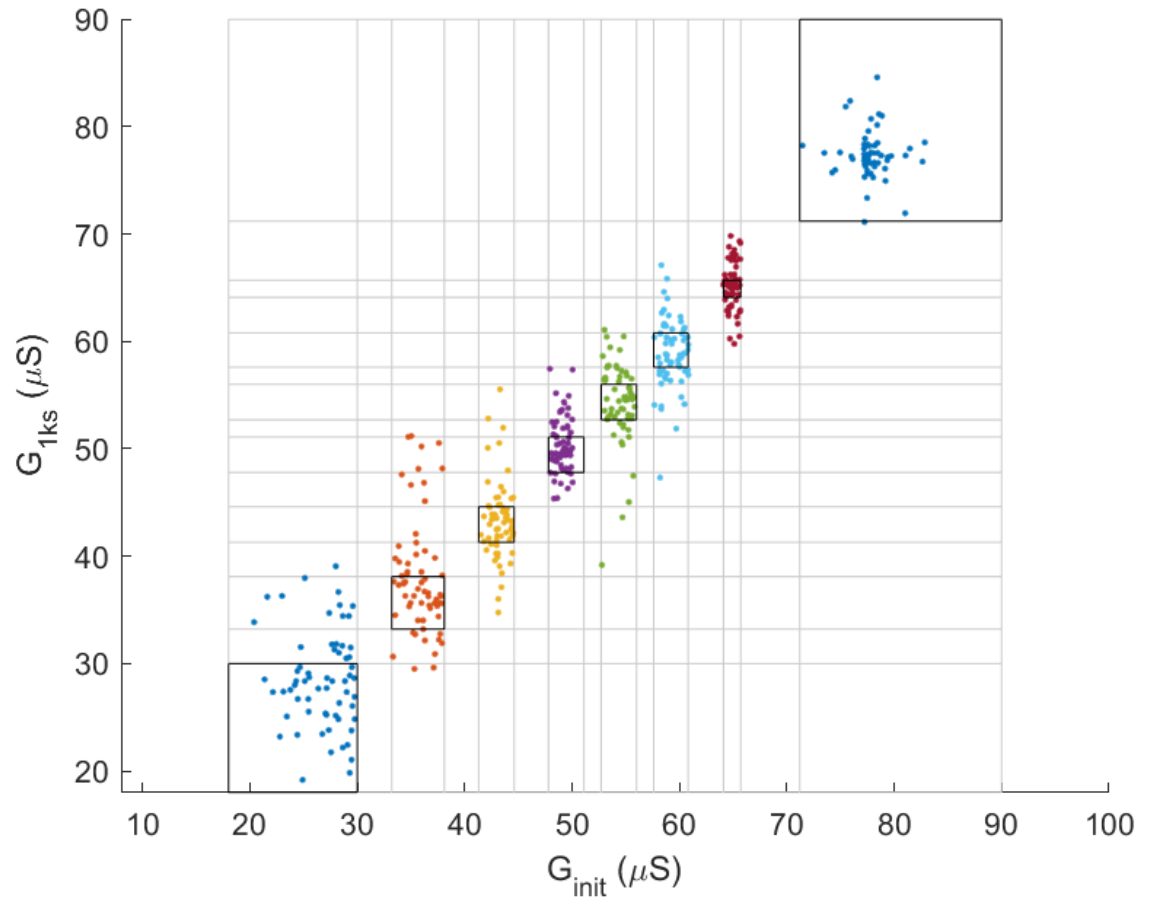

Fig. 9. Final conductance values for the 64 devices, measured 1ks after programming when using the slower short-term relaxation-aware programming technique in Fig. 6.

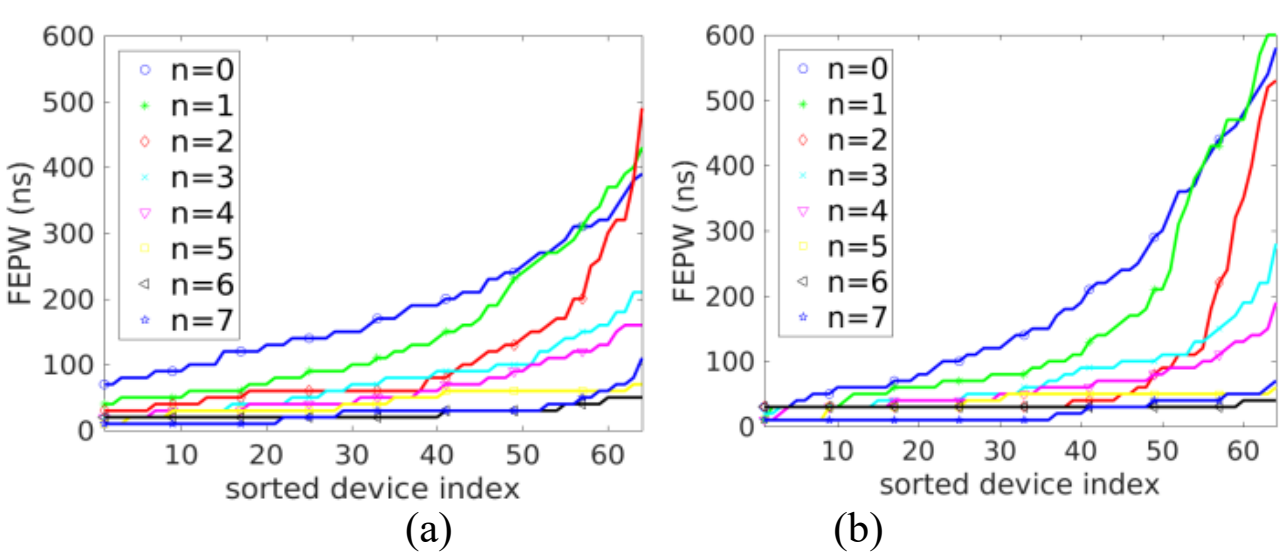

Fig. 10.The graph shows the final erase pulse width used for each of the 64 1T1R devices and for each programmed state. The 1T1R indexes are sorted by increased pulse width for each state. (a) Results for the faster algorithm in Fig. 4. (b) Results for the slower short-term relaxation-aware algorithm in Fig. 6.

## IV. Conclusions

In this paper we propose a fully digital short-term relaxation-aware erase pulse width multi-level programming technique for standard HfOx-based 1T1R crossbars. We report experimental measurements on a 64-device crossbar fabricated in CEA-LETI monolithic CMOS+RRAM 130nm technology. The experimental results demonstrate that it is possible to obtain 4-level (2-bit) programming, verified 1000 seconds after full array programming. Table IV compares with other relaxation-aware reported methods that use CC (current compliance), PAM (pulse amplitude modulation), or PWM (pulse width modulation) to perform multi-level programming.